\documentclass[12pt,a4paper]{article}
\usepackage{graphicx}
\usepackage[cp1251]{inputenc}
\usepackage[english]{babel}
\usepackage{amssymb, amsfonts, amsmath}
\usepackage{amscd}
\begin{document}

\title{Algebra of symmetry operators for Klein-Gordon-Fock equation}

\author{V.V. Obukhov}

\date{}

\maketitle

Tomsk State Pedagogical University, 60 Kievskaya St., Tomsk, 634041, Russia.

 e.mail: obukhov@tspu.edu.ru.

Tomsk State University of Control Systems and Radio Electronics,36, Lenin Avenue, Tomsk, 634050, Russia

\abstract

All external electromagnetic fields in which the Klein-Gordon-Fock equation admits the first-order symmetry operators are found, provided that in the space-time \quad $V_4$\quad a  group of motion \quad $G_3$ \quad acts simply transitively on a non-null subspace of transitivity  \quad $V_3$. \quad It is shown that in the case of a Riemannian space \quad $V_n$,\quad in which the group \quad $G_r$\quad acts simply transitively, the algebra of symmetry operators of the $n$-dimensional Klein-Gordon-Fock equation in an external admissible electromagnetic field coincides with the algebra of operators of the group\quad $G_r$.

\quad


\section{Introduction}

The Klein-Gordon-Fock equation describes the dynamics of a charged massive scalar particle interacting with an electromagnetic field. Recently, interest in the Klein-Gordon-Fock equation has grown significantly due to attempts to solve the problem of dark matter in the framework of the scalar-tensor theory.
For the successful construction of a realistic model, it is especially important to have, if not an exact solution of the basic equation, then at least a reliable approximate solution. This possibility was provided by the method of complete separation of variables, which makes it possible to reduce the original equations in partial variables to systems of ordinary differential equations. In this case, obtaining approximate solutions is not a problem, while traditional approaches in the framework of perturbation theory may turn out to be ineffective.
We should note that when finding all known exact solutions of the gravitational field equations (including self-consistent ones) on certain stages, there is always a separation of variables.
The symmetry of the classical and quantum one-particle motion equations is closely related to the symmetry of the space itself. If space admits sets of geometric objects consisting of the Killing fields, then the motion equations also have symmetry operators defined by the same sets. From a physical point of view, sets of three geometric objects are of particular interest. In the Stackel spaces, such sets are called complete.
They consist of mutually commuting Killing vector and tensor fields. Homogeneous spaces are other interesting types of spaces with three geometric objects. In them, the group of motion  \quad $G_3$ \quad acts simply transitively on a non-null hypersurface of transitivity. The theory of complete separation of variables (or the theory of the Stackel spaces) is a consequence of the theory of symmetry. The complete separation of variables in the classical, and under certain conditions, in the quantum motion equations for a test particle is possible only in the Stackel space. The Stackel spaces were named after Paul Stackel, who gave the first example of such space \quad \cite{1}. \quad Besides Stackel, useful and applicable contributions to the construction of the theory were made by  Levi-Civita \quad \cite{2},\quad  Yarov-Yarovoy \quad \cite{3}, \quad and Shapovalov \quad\cite{4}-\cite{6}.\quad  V. V. Shapovalov proved the main theorem of the theory of Stackel spaces. The theorem makes it possible to carry out an invariant partition of the of the Stackel  metrics set into equivalence classes. This made it possible to solve classification problems separately for each type of space (to list all nonequivalent metrics and electromagnetic potentials).

In paper \quad \cite{7}, \quad the theory was generalized to the case of complex privileged coordinate systems. A more detailed description of the theory and a fairly detailed bibliography can be found in the works \quad \cite{8}-\cite{10}.\quad Shapovalov's theorem is of great theoretical and applied relevance. In particular, the theorem made it possible to obtain and systematize all cases of complete separation of variables in one-particle equations of classical and quantum mechanics in flat space-time.
A large number of papers have been devoted to the theory of complete separation of variables since its inception. Nevertheless, it can be considered complete only for the free Hamilton-Jacobi equation. After the publication of articles \quad \cite {10} - \cite {13} \quad the classification problem of complete separation of variables for the Hamilton-Jacobi equation in an external electromagnetic field has been solved.
However, even for the free Klein-Gordon-Fock equation, the problem of constructing and classifying the Stackel spaces is still topical, since the equations that define, according to Shapovalov's theorem, the necessary and sufficient conditions for the complete separation of variables have not yet been solved in a general case. Moreover, they have not been solved for the free Dirac-Fock equation, as well as for all quantum equations of motion in the external fields of a gauge nature. Only isolated results have been obtained. For example, the problem of complete separation of variables in the Klein-Gordon-Fock equation has been solved for the Einstein spaces and for vacuum solutions of the Einstein equations (see.\quad \cite{14}-\cite{17}).\quad
In the papers \quad \cite{18}-\cite{20} \quad intersections of  sets of homogeneous spaces and the Stackel spaces have been considered.
Due to the high level of symmetry of the Stackel spaces, they remain interesting objects for research in various branches of theoretical and mathematical physics. An important direction is associated with the study of geometry and physics in spaces and fields that admit complete separation of variables in quantum equations of motion (see for, example, \quad \cite{21} - \cite{25}).

Solutions of scalar equations are widely used in cosmology, including in the study of the  of dark matter and dark energy problem (see, for example, \quad \cite {26}.) \quad Methods of the symmetry theory of are used to justify the choice of models of the  extended gravity cosmology (see, for example, \quad \cite {27}), \quad  and to find realistic models  \quad - \quad \cite {28}, \quad \cite {28a},\quad  etc.

Let us note one more feature of the Stackel spaces. The presence vector and tensor Killing fields in space-time allows  separating  variables in the Einstein equation, since in a privileged coordinate system the metric contains only functions, arbitrarily depending on one of the non-privileged variables. The fact that the metric is given with the indicated arbitrariness makes it possible to reduce the problem of integrating field equations and equations of motion to the problem of solving functional equations and, as a consequence, to the problem of integrating systems of ordinary differential equations.
The same arbitrary nature is peculiar for space-time manifold \quad $V_4$, \quad when a group of motions \quad $G_3$\quad acts simply transitive on a subspace \quad $V_3$. \quad

The gravitational equations for the space-time manifolds with such groups also admit the separation of variables, and the solution of the field equations is reduced to the classification of the energy-momentum tensor of matter admitted by a given geometry.
The methods of the complete separation of variables theory  in these spaces are generally inapplicable. For them, it is possible to use the method of linear partial differential equations integration, developed in \quad \cite{29},\quad  using non-commutative algebras of symmetry operators. The method made it possible to significantly expand the classification of external fields and  Riemannian manifolds that admit the existence of exact solutions of the Klein-Gordon-Fock equation, and served as the foundation for the study of quantum effects in homogeneous spaces.
In particular, in \quad  \cite{30}-\cite{32},\quad  a complete classification of spaces admitting a simply transitive action of the motions groups \quad $G_4$\quad was obtained, provided that the Klein-Gordon-Fock equation is exactly solved by non-commutative integration methods. In \quad \cite{33} - \cite{37},\quad  a similar problem was solved for Dirac-Fock equation.

In this article, a complete classification of admissible electromagnetic fields is carried out for the case when the groups of motions \quad $ G_3 $ \quad act simply transitively on the nonzero subspace \quad $ V_3 $ \quad of the space-time manifold \quad $ V_4 $. \quad All the corresponding electromagnetic fields for such groups are found.  By admissible we mean fields for which the Klein-Gordon-Fock equation admits symmetry operators.

Let us note the following circumstance.
As it was already noted, all external electromagnetic fields in which the Hamilton-Jacobi equation admits complete separation of variables for the test charge were found \quad \cite{10}-\cite{13}.\quad Thus, the problem under consideration can be viewed as an extension of the work, the final goal of which is to classify  all admissible external electromagnetic fields, both with respect to the action of the symmetry operators of the complete set and with respect to the action of the operators of the group\quad $G_r$. \quad

\maketitle

\section{Conditions for the existence of symmetry operators}

Consider a  Riemannian space \quad $ V_n $.\quad in which a group \quad $ G_r $ \quad  acts simply transitively on a subspace \quad $ V_r $.\quad
Coordinate indexes of variables in the local coordinate system \quad $[u^i] $\quad of the space \quad $V_n$\quad will be denoted as follows: \quad $i, j, k, l =0, 1,\dots n-1.$
The transitivity subspace where the group \quad $ G_r $\quad acts is given by the system of equations:
$$\psi^p(u^i)=const, \quad (p, q =0,\dots , n-r-1).$$
In what follows, it is assumed that \quad $\psi^p=u^p.$ \quad Then the transitivity surface is given by the condition \quad $ u^p=const.$ \quad The local coordinate system in the subspace \quad $ V_r $\quad will be denoted $[u^\alpha]$. Indexes \quad $\alpha, \beta, \gamma, \sigma, \tau $ \quad range from \quad $n-r$ \quad to \quad $n-1$ \quad

There is a summation within the specified limits of index change on repeated upper and lower indexes. The subject of our study are the conditions for the existence of the algebra of first order symmetric operators  (integrals of motion)  of the classical and quantum motion equations for a charged scalar test particle in an external electromagnetic field.

\subsection{Hamilton-Jacobi equation.}
Let us consider the Hamilton-Jacobi equation for a charged test particle in an external electromagnetic field with the potential \quad $A_i$:
\begin{equation}\label{1}
  g^{ij}P_iP_j=m, \quad P_i=p_i+A_i,\quad p_i=\frac{\partial{\varphi}}{\partial{u^i}}.
\end{equation}
    It is commonly known that the first order integrals of motion  of the free Hamilton-Jacobi equation are given by Killing vector fields \quad $\xi^i_\alpha$ \quad and have the following form
    \begin{equation}\label{2}
      Y_\alpha=\xi_\alpha^i p_i.
    \end{equation}
     Let us show that in case if the equation \quad \eqref{1}\quad has\quad $r$ \quad independent first-order integrals of motion, these integrals have the form of \quad \eqref{2}.\quad We will try to find the solution of the motion integrals in the form:

    \begin{equation}\label{3}
    \hat{Y}_\alpha=\zeta_\alpha^ip_i.
    \end{equation}
    The equation \quad \eqref{1} \quad allows the motion integrals of the form \quad \eqref{3}\quad if \quad $H$\quad and \quad $\hat{Y_\alpha}$\quad commute with respect to the Poisson brackets:

    \begin{equation}\label{4}
      [H,\hat{Y}_\alpha]_P=\frac{\partial H}{\partial p_i}\frac{\partial \hat{Y}_\alpha}{\partial x^i} - \frac{\partial H}{\partial x^i}\frac{\partial \hat{Y}_\alpha}{\partial p_i}= (g^{il}{\zeta^{j}_\alpha}_{,l}+g^{jl}{\zeta^{i}_\alpha}_{,l}-g^{ij}_{,l}\zeta_\alpha^l)P_i P_j + 2g^{i\sigma}(\xi^{j}_\alpha F_{ji}+(\zeta_{\alpha}^\beta A_\beta)_{,i})P_\sigma=0.
    \end{equation}
     The functions \quad $\zeta^{j}_\alpha$\quad satisfy the Killing equations:
     $$g^{il}{\xi^{j}_\alpha}_{,l}+g^{jl}{\xi^{i}_\alpha}_{,l}-g^{ij}_{,l}\xi_\alpha^l=0,$$
     and therefore they coincide with the Killing vector field:
      $$\zeta^{j}_\alpha=\xi^{j}_\alpha.$$
      The coefficients before\quad $P_iP_j$\quad in the equations \quad \eqref{4} \quad must vanish.
      Therefore, from the equations \quad \eqref{4}\quad it follows:

    \begin{equation}\label{5}
      (\xi_{\alpha}^j A_j)_{,i} = \xi^{j}_\alpha F_{ij}.
    \end{equation}
In contrast to the free Hamilton-Jacobi equation, the equation \quad\eqref{1} \quad in the general case has no integrals of motion. The system of equations \quad \eqref{5} \quad defines the set of admissible electromagnetic fields. In these fields, the equation \quad\eqref{1} \quad has \quad $r$ \quad first-order integrals of motion defined by the group \quad $G_r.$ \quad
The transitivity subspace where the group \quad $ G_r $\quad acts is given by the system of equations:
We consider the following subsystem of the system\quad\eqref{5}:
\begin{equation}\label{6}
    (\xi_{\alpha}^j A_j)_{,p} = \xi^{j}_\alpha F_{pj}.
\end{equation}
Since in the chosen coordinate system \quad $\xi^{j}_{\alpha,p}=0,$\quad from the equations \quad \eqref{5} \quad it follows:

    \begin{equation}\label{7}
    \xi_\alpha^\beta A_{p,\beta}=0 \quad\rightarrow \quad A_p=A_p(u^q).
    \end{equation}
Thus, the components \quad $A_p$\quad can be made zero by the gradient transformation of the potential. Further, we will select the calibration of the potential in exactly this way:
    $$A_p=0.$$
      Then from \quad \eqref{5}\quad it follows:
\begin{equation}\label{7a}
(\xi_{\alpha}^\beta A_\beta)_{,\gamma} = \xi^{\beta}_\alpha F_{\gamma\beta}.
\end{equation}
 We prove, that \quad \eqref{7a} \quad is compatible. Indeed the  system \quad \eqref{7a} \quad can be present in the form:

\begin{equation}\label{7b}
\hat{Y}_\alpha \textbf{A}_\beta = C^\gamma_{\alpha\beta}\textbf{A}_\gamma, \quad  \quad \textbf{A}_\beta = \xi^\alpha_\beta A_\alpha,
\end{equation}
Then the compatibility conditions can be transformed as follows:
$$
(\delta^\sigma_\alpha[\hat{Y}_\beta \hat{Y}_\gamma] - C^\sigma_{\gamma\alpha}\hat{Y}_\beta + C^\sigma_{\beta\alpha} \hat{Y}_\gamma)\textbf{A}_\sigma=(C^\tau_{\alpha\beta}C^\sigma_{\gamma\tau} + C^\tau_{\beta\gamma}C^\sigma_{\alpha\tau} +C^\tau_{\gamma\alpha}C^\sigma_{\beta\tau})\textbf{A}_\sigma = 0.
$$
(The Bianchi identity is used.)

In \quad \cite{32} \quad it was proved that the Hamilton-Jacobi and the Klein-Gordon-Fock equations admit a symmetry operator of the form
\begin{equation}\label{8}
 \hat{Y}_\alpha=\xi_\alpha^i(\hat{p}_i +A_i) + \gamma_\alpha.
 \end{equation}
if and only if the electromagnetic field satisfies the system of equations
\begin{equation}\label{9}
\gamma_{\alpha,i} = \xi^{j}_\alpha F_{ji}
\end{equation}
	By comparing the equations \quad\eqref{5}\quad and \quad \eqref{6}\quad one can show that:
 \begin{equation}\label{10}
 \gamma_\alpha = -\xi_\alpha^\beta A_\beta.
 \end{equation}
 Therefore the operator \quad \eqref{8} \quad takes the form \quad \eqref{2}.

\subsection{Klein-Gordon-Fock equation}

Let us consider the Klein-Gordon-Fock equation:
\begin{equation}\label{11}\hat{H}\varphi=(g^{ij}\hat{P}_i\hat{P}_j)\varphi = m\varphi, \quad \hat{P}_j = -\imath \hat{\nabla}_i + A_i.\end{equation}
$\hat{\nabla}_i$ - operator of the covariant derivative, with metric-compatible connectivity, corresponding to the operator of the partial derivative\quad - \quad $\hat{\partial}_i = \imath \hat{p}_i$ \quad with respect to the coordinate \quad $u_i$;\quad $\varphi$ \quad is a field of a scalar particle with mass \quad $m.$

We denote the Laplase-Beltrami operator as \quad $\hat{H}_0$:
$$\hat{H}_0=g^{ij}\hat{\nabla}_i\hat{\nabla}_j.$$
Then the  operator \quad $\hat{H}$ \quad  can be presented as:
\begin{equation}\label{12}
\hat{H} = -\hat{H}_0 + \hat{H}_e. \quad \hat{H}_e = 2A^i\hat{p}_i - \imath(\hat{\nabla}_iA^i) + A_i A^i.
\end{equation}
Let us prove the

\quad

 \textbf{Statement}.

\quad

 Algebra of symmetry operators of the Klein-Gordon-Fock equation
coincides with the Lie algebra of the group \quad $G_r$.\quad That is   the Klein-Gordon-Fock operator commutes with the operators
\begin{equation}\label{13}
\hat{Y}_\alpha=\xi^i_\alpha\hat{p}_i,
\end{equation}
for the admissible external electromagnetic field.

\quad

Indeed the  commutator of operators \quad $\hat{H}$\quad and \quad $\hat{p}_i$ \quad has the form:
$$
[\hat{H},\hat{Y}_\alpha^i]=[\hat{H}_0,\hat{Y}_\alpha^i] + [\hat{H}_e,\hat{Y}_\alpha^i]=0.
$$
As it is known,

 \begin{equation}\label{14}
 [\hat{H}_0,\hat{Y}_\alpha^i] = 0 \rightarrow  [g^{ij}\hat{\partial}_i \hat{\partial}_j+(g^{ij}_{,j}+g^{ij}\chi_{,j}\hat{\partial}_i),\xi_\alpha^i \hat{\partial}_i]=0.\end{equation}
We denoted here: $\quad 2\chi_i = -g_{kl}g^{kl}_{,i}$.\quad Since the functions \quad $\xi_\alpha^i$ \quad satisfy the Killing equations, from \quad \eqref{14} \quad  we get the condition:
\begin{equation}\label{15}
g^{il}{\xi_\alpha^k}_{,il}+(g^{il}_{,l}+g^{il}\chi_{,l}){\xi_\alpha}^k_{,i} - (g^{il}_{,l}+g^{il}\chi_{,l})_{,i} \xi_\alpha^i=0.
\end{equation}
Using the consequences from the Killing equations we get:
$$g^{il}_{,il}{\xi_\alpha^k} = g^{ik}{\xi_\alpha^l}_{,il} + g^{il}_{,l}{\xi_\alpha^k}_{,i},$$
$$ g^{il}{\chi_l}{\xi_\alpha^k}_{,i} - {(g^{il}{\chi_{,l}})}_{,i}{\xi_\alpha^i} = g^{il}(\chi_{,l}{\xi_\alpha}^i)_{,i.}$$
It allows to reduce \quad\eqref{15} \quad  to the form:
$$({\xi_\alpha^i}_{,i}+\chi_i\xi_\alpha^i)_{,l}=0.$$
Since:
$$2({\xi_\alpha^i}_{,i}+\chi_i\xi_\alpha^i)={\xi_\alpha}^i g_{kl}g^{kl}_{,i}+2{\xi_\alpha^i}_{,i}=0,$$
the equation \quad \eqref{15} \quad is reduced to an identity.
Because the commutator \quad $[\hat{H}_e,\hat{Y}_\alpha]\quad $ equals to zero  we obtain necessary and sufficient conditions for the existence of the symmetry operators algebra of the equation \quad \eqref{11} \quad in the form:
\begin{equation}\label{16}
A^l{\xi_\alpha}^k_{,l}-A^k_{,l}\xi_\alpha^l = 0 \quad \rightarrow (\xi_\alpha^\beta A_\beta)_{,i}=\xi_\alpha^\beta F_{i \alpha}.
\end{equation}
\begin{equation}\label{17}
  \xi_\alpha^i (A_lA^l)_{,i}=0.
\end{equation}
\begin{equation}\label{18}
{\xi_\alpha}^k(A^{l}_{,l}+A^{l}\chi_{,l})_{,k}=0
\end{equation}
The equation \quad \eqref{17} \quad is a consequence of the equations \quad \eqref{16} \quad and the Killing equations, because:
$$\quad {\xi_\alpha}^i A_l{A^l}_{,i}+{\xi_\alpha}^i A^l{A_{l,i}}=A_lA^k{\xi_\alpha^l}_{,k}+{\xi_\alpha}^i A^l{A_{l,i}}=(\xi_\alpha^\beta A_\beta)_{,i}+\xi_\alpha^\beta F_{\beta i}=0.$$
Let us consider the equation \quad \eqref{18}.\quad Using the condition \quad \eqref{16} \quad the first term can be presented as:
$$ {\xi^k_\alpha}{A^l}_{,kl}=({\xi^k_\alpha}{A^l}_{,k})_{,l}-\xi^k_{\alpha,l}{A^l}_{,k}= (A^k{\xi^l_{\alpha,kl}}+{\xi_{\alpha,l}^k}{A^l}_{,k})-\xi^k_{\alpha,l}{A^l}_{,k}=A^k{\xi^l_{\alpha,kl}}$$
Second term we transform as following:
$$(A^l\chi_l)_{k}\xi_\alpha^k=A^k(\xi_{\alpha,k}^l\chi_l+\xi_{\alpha}^l\chi_{,lk}) \quad \rightarrow \quad {A}^k(\xi^{l}_{\alpha,l}+\xi^{l}_\alpha\chi_{,l})_{,k}=0.$$
Thus only one condition \eqref{16} remains. The algebra exists, if and only if the admissible electromagnetic field   exists.

\quad

The \textbf{Statement} is proved.

\quad
\quad

The integrals of motion of the Hamilton-Jacobi and the Klein-Gordon-Fock equations exist for the same electromagnetic fields and have the same form. In order to find the admissible electromagnetic fields, it is necessary to investigate the compatibility conditions of the system \quad \eqref{6} \quad having the form:
\begin{equation}\label{19}
{\gamma_\sigma}_{,\alpha i}-{\gamma_\sigma}_{,i\alpha} = \xi_\sigma^\beta F_{i\alpha,\beta}+\xi_{\sigma,i}^\beta F_{\beta \alpha}-\xi_{\sigma,\alpha}^\beta F_{\beta i}=0.
\end{equation}
The system \quad \eqref{19} \quad must be supplemented with the Bianchi identities.
One can  use the solutions \quad $F_{i \alpha}$ \quad  to find the potential of the admissible electromagnetic field by integrating the compatible system of equations:
$$
  A_{\alpha,\beta}-A_{\beta,\alpha}=F_{\beta\alpha}.
$$
We follow  A. Z. Petrov \quad  \cite{38}. \quad This book contains all metrics of space-time manifolds in which the groups \quad $G_3(N)$ \quad act. We will hold to the notation accepted in this book with exception - the non-ignored variables will be denoted by\quad  $u^0$. \quad In addition for convenience we will use the notation \quad $u^i=u_i.$ \quad  Functions that depend only one the variables
\quad $u^0 = u_0$\quad are denoted by by lowercase Greek letters with a single right subscript.\quad Examples: \quad $\alpha_\alpha =\alpha_\alpha(u^0);\quad \xi^\alpha_\beta= \xi^\alpha_\beta (u^i). $ Constants are denoted by by lowercase Latin letters with the $tilda$ symbol. Derivatives with respect to the  variables $u_0$ are denoted by dots. Example: \quad $\dot{\alpha_0}=\partial\alpha/\partial u_0. \quad$

\section{Solvable groups $G_3(N).$}

	According to the Bianchi classification, depending on the set of structural constants \quad $C^\gamma_{\alpha\beta}$, \quad there are 9 types of groups\quad $G_3(N).$\quad The first seven groups are solvable. Let us list them.

    Solvable groups:
    \begin{equation}\label{20}
    \left\{\begin{array}{ll}
    G_3(I):\quad C^\gamma_{\alpha\beta}=0 ;\quad \cr
    G_3(II):\quad C^\alpha_{12}=0,\quad C^\alpha_{13}=0 \quad C^\alpha_{23}= \delta^\alpha _1;\quad \cr
    G_3(III):\quad C^\alpha_{12}=0, \quad C^\alpha_{13}=\delta^\alpha_1 \quad C^\alpha_{23}=0;\quad \cr
    G_3(IV):\quad C^\alpha_{12}=0, \quad C^\alpha_{13}=\delta^\alpha _1\quad C^\alpha_{23}=
    \delta^\alpha _1 + \delta^\alpha_2;\quad \cr
    G_3(V):\quad C^\alpha_{12}=0, \quad C^\alpha_{13}=\delta^\alpha_1\quad C^\alpha_{23}=
    \delta^\alpha _2;\quad \cr
    G_3(VI):\quad C^\alpha_{12}=0, \quad C^\alpha_{13}=\delta^\alpha_1,\quad C^\alpha_{23}=
    q\delta^\alpha_2.\quad (q\ne 0, 1);\quad \cr
    G_3(VII):\quad C^\alpha_{12}=0, \quad C^\alpha_{13}=\delta^\alpha_1\quad C^\alpha_{23}=
    2\delta^\alpha_2 \cos{\alpha},\quad \alpha=const.
    \\\end{array}\ \right.\end{equation}
The groups \quad $G_3(I)-G_3(VII)$ \quad contain the Abelian subgroup with the Killings vectors \quad $\xi_1^i=\delta_1^i, \quad \xi_2^i=\delta_2^i.\quad $ Therefore from \quad \eqref{19}\quad and from the Bianchi identities, it follows:

\begin{equation}\label{21}
  F_{ij,1}=F_{ij,2}=F_{12,3}=F_{12,0}=0,\quad F_{13,0}+F_{01,3}=0,\quad
F_{23,0}+F_{02,3}=0.\end{equation}
For the groups \quad $G(I) - G(VI)$ \quad the functions \quad $\xi_3^\alpha, \quad \xi_{3,i}^\alpha$ \quad can be presented as the general formula:
\begin{equation}\label{22}
  \xi_3^\alpha = (ku_1 +\varepsilon u_2)\delta_1^\alpha + n u_2 \delta_2^\alpha - \delta_3^\alpha,\quad  \xi_{3,i}^\alpha = (k\delta_{i1} +\varepsilon \delta_{i2})\delta_2^\alpha + n \delta_{i2}\delta_2^i.
\end{equation}
Parameters: \quad $k,\quad \varepsilon, \quad n\quad $ for each number \quad $N$ \quad take values:

 \quad $N=I\quad \quad  \rightarrow k=n=\varepsilon=0,$

 \quad $N=II\quad  \rightarrow k=n=0,\quad \varepsilon=1,$

 \quad $N=III\quad \rightarrow k=1, \quad n=\varepsilon=0,$

 \quad $N=IV\quad  \rightarrow k=n=\varepsilon=1,$

 \quad $N=V\quad \quad \rightarrow k=n=1, \quad\varepsilon=0,$

 \quad $N=VI\quad \rightarrow k=1, \quad n=2,\quad\varepsilon=0.$\quad

The equation system \quad \eqref{19} \quad together with the Bianchi identities has the form:

\begin{equation}\label{23}
\left\{\begin{array}{ll}
F_{01,3}=k F_{01},\quad F_{02,3}=\varepsilon F_{01}+nF_{02},\quad  F_{03,3}=0, \cr
F_{13,3}=k F_{13},\quad  F_{12,3}=(k+n)F_{12},\quad F_{23,3}=\varepsilon F_{13}+nF_{23},\cr
F_{12,0}=F_{12,3}=0,\quad F_{01,3}+F_{13,0}=0,\quad  F_{02,3}+F_{23,0}=0.
\\\end{array}\ \right.
\end{equation}
For the group \quad $G(VII)$ \quad the functions \quad $\xi_3^\alpha, \quad \xi_{3,i}^\alpha$ \quad can be presented as following:
 \begin{equation}\label{24}
   \xi_3^\alpha=-u_2\delta^\alpha_1 +(2u_2 \cos{\alpha}+u_1)\delta^\alpha_2+\delta^\alpha_3
   \quad \xi_{3,i}^\alpha=-\delta_{i2}\delta^\alpha_1 +(2\delta_{i2} \cos{\alpha}+\delta_{i1})\delta^\alpha_2.
 \end{equation}
 The formulas \quad \eqref{24} \quad differ from ones listed in the book \quad \cite{38} \quad  in regard to the exchange:
 $$
 q=2\cos{\alpha}, \quad (\alpha=const).
 $$
  Let us consider the equations \quad\eqref{23}\quad for each of the groups $G_3(I)-G_3(VI).$

\quad

\subsection{ The group \quad $G_3(I).$}

\quad

 In this case\quad $\xi_3^i=\delta_3^i.$ \quad  From \quad \eqref{23} \quad it follows:
 \begin{equation}\label{25}
 F_{ij,3}=0,\quad F_{\alpha\beta,0}=0 \rightarrow  F_{\alpha0}=\dot{\beta _\alpha}(u_0).
 \end{equation}
Potential $A_i$ is derived from the system of differential equation:

 \begin{equation}\label{26}
 A_{j,i}-A_{j,i}=F_{ij},
 \end{equation}
and has the form:
$$ A_\alpha = \beta_\alpha + \tilde{c}_{\alpha \beta}u^\beta, \quad \tilde{c}_{\alpha \beta}=-\tilde{c}_{ \beta \alpha}$$
Using the equations \quad \eqref{7b}, \quad one can show that\quad $\tilde{c}_{\alpha \beta}=0$. \quad

\subsection{ Group \quad $G_3(II).$}

In this case \quad $k=n=0,\quad \varepsilon=1$ \quad and the system of equation \quad \eqref{23}\quad  have the form:
\begin{equation}\label{27}
\left\{\begin{array}{ll}
  F_{12,3}=F_{12,0}= F_{13,3}=F_{03,3}=0;\cr
  F_{02,3}=F_{01},\quad F_{23,0}+F_{02,3}=0,\quad F_{23,3}+F_{13}=0;\cr
  F_{13,0}+F_{01,3}=0,\quad  F_{23,0}+F_{02,3}=0.
  \\\end{array}\ \right.
 \end{equation}
From the first equations if the system \quad \eqref{27} \quad  it follows:
$$ F_{12}=\tilde{a},\quad F_{03}=\dot{\gamma_0}, \quad F_{13}=2\tilde{b}.$$
By placing these equations in the remaining equations of the system \eqref{28} we get:
\begin{equation}\label{28}
F_{01}=\dot{\alpha}_0,\quad F_{02}=\dot{\alpha}_0 u_3+\dot{\beta}_0, \quad F_{23}=-\alpha_0 +2\tilde{b} u_3.
\end{equation}
A particular solution of the system \eqref{26} in the selected gauge has the form:
\begin{equation}\label{29}
A_1=\alpha_0-2\tilde{b}u_3,\quad A_2=\alpha_0 u_3+\beta_0 +\tilde{a}u_1 -\tilde{b}u_3 ^2, \quad A_3=\gamma_0.
\end{equation}
Using the equations \quad \eqref{7b}, \quad one can show that\quad $ \tilde{a}=\tilde{b}=0$. \quad

\subsection{ Group \quad $G_3(III).$}

In this case \quad $k=0,\quad n=\varepsilon=0$ \quad and the systems of equations \quad \eqref{23}\quad have the form:

 \begin{equation}\label{30}
\left\{\begin{array}{ll}
F_{03,3}=F_{02,3}=F_{23,3}=F_{12,3}=F_{12,0}=0;\cr
 F_{01,3}=F_{01},\quad F_{12,3}=F_{12}, \quad F_{13,3}=F_{13}; \cr
 F_{23,0}+F_{02,3}=0, \quad F_{13,0}+F_{01,3}=0,.
  \\\end{array}\ \right.
 \end{equation}
Hence we find the functions \quad $F_{ij}$:
$$
F_{01}=\dot{\alpha_0}\exp{u_3},\quad F_{02}=\dot{\beta_0},\quad
F_{03}=\dot{\gamma_0},
$$
$$
\quad F_{12}=0, \quad F_{13}=-\alpha_0\exp{u_3},\quad F_{23}=\tilde{a}.
$$
A particular solution of the system \quad \eqref{26} \quad in the selected gauge has the form:

\begin{equation}\label{31}
 A_{1}=\alpha_0\exp{u_3},\quad A_2=\beta_0,\quad A_3=\gamma_0+\tilde{a}u_2.
\end{equation}
Using the equations \quad \eqref{7b}, \quad one can show that\quad $ \tilde{a}=0$. \quad

\subsection{ Group \quad $G_3(IV).$}

In this case \quad $k=n=\varepsilon=1$ \quad and the systems of equations \quad  \eqref{23} \quad  have the form:
 \begin{equation}\label{32}
\left\{\begin{array}{ll}
F_{01,3}=F_{01},\quad F_{02,3} = F_{01}+F_{02},\quad F_{03,3} = 0,\cr
F_{12,3} =2F_{12}, \quad F_{13,3}=F_{13},\quad
 F_{23,3}=F_{23}+ F_{13}=0,\cr
F_{13,0}+F_{01,3}=0,\quad F_{23,0}+F_{02,3}=0,\quad F_{01,3}+F_{13,0}=0, \quad F_{12,3}=0.
  \\\end{array}\ \right.
 \end{equation}

From here it follows:
$$F_{01}=\dot{\alpha_0}\exp{u_3},\quad F_{02}=(\dot{\alpha_0}u_3+\dot{\beta_0})\exp{u_3},\quad F_{03}=\dot{\gamma}_0,
$$
$$ F_{12}=0, \quad F_{13}=-\alpha_0\exp{u_3},\quad F_{23}=-(\alpha_0 u^3+\alpha_0+\beta_0)\exp{u_3}.$$

A particular solution of the system \quad \eqref{26} \quad in the selected gauge has the form:

\begin{equation}\label{33}
A_1=\alpha_0\exp{u_3},\quad A_2 = (\alpha_0 u_3 + \beta_0)\exp{u_3},\quad A_3=\gamma_0.
\end{equation}

\subsection{ Group \quad $G_3(V).$}
In this case \quad $k=n=1,\quad \varepsilon=0,$\quad and the systems of equations \quad \eqref{23} \quad have the form:
 \begin{equation}\label{34}
\left\{\begin{array}{ll}
F_{01,3}=F_{01},, \quad F_{02,3} =F_{02},\quad F_{03,3} = 0;\cr
F_{12,3} = F_{12}, \quad F_{13,3}=F_{13},\quad F_{23,3}=F_{23};\cr
F_{13,0}+F_{01,3}=0,\quad F_{23,0}+F_{02,3}=0,\quad F_{12,3}=F_{12,0}=0.
\\\end{array}\ \right.
 \end{equation}
From here it follows:
$$F_{01}=\dot{\alpha_0}\exp{u_3},\quad F_{02}=\dot{\beta_0}\exp{u_3},\quad F_{03}=\dot{\gamma_0},$$
$$ F_{12}=0,\quad F_{13}=-\alpha_0\exp{u_3},\quad F_{23}=-\beta_0\exp{u_3}.$$
A particular solution of the system \quad  \eqref{26} \quad  in the selected gauge has the form:
\begin{equation}\label{35}
A_1=\alpha_0\exp{u_3},\quad A_2 = \beta_0 \exp{u_3},\quad A_3=\gamma_0.
\end{equation}

\subsection{ Group \quad $G_3(VI).$}

In this case  \quad $k=1, \quad n=2,\quad\varepsilon=0$\quad and the systems of equations \quad  \eqref{23} \quad  have the form:
\begin{equation}\label{36}
\left\{\begin{array}{ll}

 F_{01,3}=F_{01}, \quad F_{02,3} =2F_{02},\quad F_{03,3} = 0;\cr
 F_{12,3} = 3F_{12},\quad F_{13,3}=F_{13},\quad F_{23,3}=2F_{23}; \cr
  F_{12,0}=F_{12,3}=0, \quad F_{13,0}+F_{01,3}=0,\quad F_{23,0}+F_{02,3}=0,.
\\\end{array}\ \right.
 \end{equation}
From here it follows:
$$F_{01}=\dot{\alpha_0}\exp{u_3},\quad F_{02}=\dot{\beta_0}\exp{2u_3},\quad F_{03}=\dot{\gamma_0},$$
$$ F_{12}=0,\quad F_{13}=-\alpha_0\exp{u_3},\quad F_{23}=-2\beta_0\exp{2u_3}.$$
A particular solution of the system \quad \eqref{24} \quad in the selected gauge has the form:
\begin{equation}\label{37}
A_1=\alpha_0\exp{u_3},\quad A_2 = \beta_0 \exp{2u_3},\quad A_3=\gamma_0.
\end{equation}

\subsection{ Group \quad $G_3(VII).$}

In this case the relations \quad  \eqref{23}\quad  occur, and the systems of equations \quad  \eqref{19} \quad can be presented as:
$$
F_{i\alpha,3}+\delta_{1i}F_{2\alpha}-\delta_{1\alpha}F_{2i}+ \delta_{2i}(2F_{2\alpha}\cos{\alpha}-F_{1\alpha})-
\delta_{2\alpha}(2F_{2i}\cos{\alpha}-F_{1i})=0.
$$
Hence, using the Bianchi identities as well, we obtain the following system of equations:
\begin{equation}\label{38}
\left\{\begin{array}{ll}
F_{01,3}+F_{02}=0,\quad F_{02,3}+2cos{\alpha}F_{02}-F_{01}=0,\quad  F_{03,3}=0;\cr
F_{12} = 0,\quad F_{13,3}+F_{23}=0,\quad F_{23,3}+2cos{\alpha}F_{23}-F_{13}=0, \cr
F_{13,0}+F_{01,3}=0,\quad F_{23,0}+F_{02,3}=0,.
\\\end{array}\ \right. \end{equation}
First of all, let us find the functions \quad $F_{13},\quad F_{02}:$

$$F_{13}=(\nu_0 \sin{(u_3\sin{\alpha})} +\mu_0 \cos{(u_3\cos{\alpha})})\exp{(-u_3 \cos{\alpha})},$$
$$F_{02}=(\dot{\alpha_0} \sin{(u_3\sin{\alpha})} +\dot{\beta_0} \cos{(u_3\cos{\alpha})})\exp{(-u_3 \cos{\alpha})},$$
By placing them into the system's equations \quad \eqref{38},\quad  we find the relation between the functions \quad $\alpha_0, \beta_0, \nu_0, \mu_0:$
$$\nu_0=\alpha_0,\quad \mu_0=\beta_0$$
following which, from the relations:
$$F_{02}=-F_{01,3},\quad F_{23}=-F_{13,3}, \quad F_{03}=0,$$
we define the functions \quad $F_{01},\quad F_{23},\quad F_{03}:$
$$F_{23}=(\alpha_0 \sin{(-\alpha+u_3\sin{\alpha})} +\beta_0 \cos{(-\alpha+u_3\cos{\alpha})})\exp{(-u_3 \cos{\alpha})},$$
$$ F_{01}=(\dot{\alpha_0} \sin{(\alpha+u_3\sin{\alpha})} +\dot{\beta_0} \cos{(\alpha +u_3\cos{\alpha})})\exp{(-u_3 \cos{\alpha})},\quad F_{03}=\dot{\gamma_0} $$
A particular solution of the system \eqref{26} in the selected gauge has the form:
\begin{equation}\label{39}
A_{1}=(\alpha_0 \sin{(\alpha+u_3\sin{\alpha})} +\beta_0 \cos{(\alpha+u_3\cos{\alpha})})\exp{(-u_3 \cos{\alpha})},$$
$$\quad A_{2}=(\alpha_0 \sin{(u_3\sin{\alpha})} +\beta_0 \cos{(u_3\cos{\alpha})})\exp{(-u_3 \cos{\alpha})},\quad A_3=\gamma_0.
\end{equation}

\section{Insolvable groups $G_3(N).$}

Unsolvable groups \quad  $G_3(VIII)$\quad  and \quad  $G_3(IX)$ \quad do not contain the Abelian subgroups and have more complex algebraic structures:

$$ G_3(VIII):\quad C^\alpha_{12}=\delta^\alpha_1, \quad C^\alpha_{13}=2\delta^\alpha_2\quad C^\alpha_{23}=
    -\delta^\alpha_3.$$
$$ G_3(IX):\quad C^\alpha_{12}=\delta^\alpha_3, \quad C^\alpha_{13}=-\delta^\alpha_2\quad C^\alpha_{23}=
    \delta^\alpha_1.$$
For both structures the Killing vector field \quad $\xi_1^i$ \quad has the form:
     $$\xi_1^i=\delta_2^i.$$
Therefore from the system \quad  \eqref{19}\quad  when \quad $\sigma=1$ \quad  it follows:
$$ F_{ij,2}=0. $$
Taking this condition into account, we consider the remaining equations of the system.

\subsection{Group $G_3(VIII).$}

  Functions \quad $\xi^\alpha_2,\quad \xi^\alpha_3$\quad and their derivatives have the form:

\begin{equation}\label{40}
\left\{\begin{array}{ll}
    \xi^\alpha_2 = u_2\delta^\alpha_2 + \delta^\alpha_3,\quad \xi^\alpha_{2,i} = \delta_{i2}\delta^\alpha_2; \cr
   \xi^\alpha_3 = \delta^\alpha_1 \exp{u_3} + \delta^\alpha_2 u_2^2 + 2u_2\delta^\alpha_3, \quad \xi^\alpha_{3,i} = \delta_{3i}\delta^\alpha_1 \exp{u_3} + 2(u_2 \delta^\alpha_2 +   \delta^\alpha_3)\delta_{2i.}
  \\\end{array}\ \right.
 \end{equation}

  Because \quad $F_{ij,2} =0,$ \quad the system of equations \quad  \eqref{19} \quad  splits into two
\begin{equation}\label{41}
\left\{\begin{array}{ll}
F_{i\alpha,3}+(\delta_{3i}F_{1\alpha}-\delta_{3\alpha}F_{1i})+ 2(\delta_{2i}2F_{3\alpha}-\delta{2\alpha}F_{3i})\exp{-u_3}=0; \cr
F_{i\alpha,3}+\delta_{1i}F_{2\alpha}-\delta_{1\alpha}F_{2i}+ \delta_{2i}(2F_{2\alpha}\cos{\alpha}-F_{1\alpha})-
\delta_{2\alpha}(2F_{2i}\cos{\alpha}-F_{1i})=0.
\\\end{array}\ \right.
\end{equation}

Hence, using the Bianchi identities as well, we obtain the following system of equations:

 \begin{equation}\label{42}
\left\{\begin{array}{ll}
  F_{12,3}+F_{12}=0,\quad F_{23,3}+F_{23}=0, \quad F_{02,3}+F_{02}=0;\cr
F_{02,1}+2F_{03}\exp{-u_3}=0,\quad F_{03,1}+F_{01}=0;\cr
  F_{12,1}+2F_{13}\exp{-u_3}=0 \quad F_{23,1}+F_{12}=0; \cr
F_{01,3} = F_{01,1} = F_{13,3}=F_{13,1} = F_{03,3}= 0;\cr
F_{12,0}+F_{02,1}=0,\quad F_{01,3}=F_{03,1}=0,\cr
F_{02,3}+F_{23,0}=0,\quad F_{12,3,0}+F_{23,1}=0.
\\\end{array}\ \right.
 \end{equation}

By integrating this system we get:

$$F_{01}=\dot{\alpha}_0, \quad F_{02}=(\dot{\alpha_0}{u_1}^2 + 2\dot{\beta_0} u_1 +\dot{\gamma_0})exp{-u_3},\quad F_{03}=-(\dot{\alpha}u_{1}+\dot{\beta_0}),$$
$$F_{12}= 2(\alpha_0{u_1}+\beta_0)\exp{-u_3}, \quad F_{13}=-\alpha_0,\quad F_{23}=(\alpha_0{u_1}^2 + 2\beta_0 u_1 +\gamma_0)\exp{-u_3},$$

A solution of the system \quad \eqref{24} \quad  in the selected gauge has the form:
$$
A_{1}=\alpha_0,\quad A_{2}=(\alpha_0{u_1}^2 + 2\beta_0 u_1 +\gamma_0)\exp{-u_3},\quad A_3=-(\alpha_0{u_1} + \beta_0).
$$

\subsection{Group $G_3(IX).$}

  Functions \quad $\xi^\alpha_2,\quad \xi^\alpha_3$\quad and their derivatives have the form:

  \begin{equation}\label{43}
   \left\{\begin{array}{ll}
     \xi^\alpha_2 = \delta^\alpha_1 {\cos{u_2}} + (\delta^\alpha_3 - \delta^\alpha_2 {\cos{u_1}}) \frac{\sin{u_2}}{\sin{u_1}},\quad \xi^\alpha_3 = \frac{\partial{\xi^\alpha_2 }}{\partial{u_2}}; \cr
    \quad \xi^\alpha_{2,i} = \sin{u_2}(-\delta_{i2}\delta^\alpha_1 +\frac{\delta_{i1}}{\sin^2{u_1}}(\delta^\alpha_2 - \delta^\alpha_3 \cos{u_1}))+
(\delta^\alpha_3 - \delta^\alpha_2 {\cos{u_1}})\delta_{i2}\frac{\cos{u_2}}{\sin{u_1}}
 \\\end{array}\ \right.\end{equation}

	Because \quad $F_{ij,2} =0,$ \quad the system of equations \eqref{19} splits into two subsystems:

\begin{equation}\label{44}
   \left\{\begin{array}{ll}
 F_{i\alpha,1}\sin{u_1}+\delta_{2i}(F_{3\alpha}-F_{2\alpha}\cos{u_1}) - \delta_{2\alpha}(F_{3i}-F_{2i}\cos{u_1})=0;\cr
F_{i\alpha,3}\sin{u_1}+(\delta_{2\alpha}F_{1i}-\delta_{2i}F_{1\alpha})\sin^2{u_1}+ \delta_{1i}(F_{2\alpha}-F_{3\alpha}\cos{u_1}) - \delta_{1\alpha}(F_{2i}-F_{3i}\cos{u_1})=0;
  \\\end{array}\ \right.\end{equation}
Hence, using the Bianchi identities as well, we obtain the following systems of equations:

\begin{equation}\label{45}
   \left\{\begin{array}{ll}

F_{01,1}= F_{03,1}= F_{13,1}=F_{03,3}=0,\cr

F_{01,3}\sin{u_1} + F_{02}-\cos{u_1}F_{03}=0;\quad \cr

F_{02,3}-F_{01}\sin{u_1}=0;\quad \cr

F_{12,3}\sin{u_1} + F_{23}\cos{u_1}=0;\quad \cr

F_{13,3}\sin{u_1} + F_{23}=0;\quad \cr

F_{23,3}-F_{13}\sin{u_1}=0.\quad

\\\end{array}\ \right.\end{equation}

\begin{equation}\label{46}
   \left\{\begin{array}{ll}
F_{02,1}\sin{u_1}-\cos{u_1}F_{02} +F_{03}=0;\quad \cr
F_{12,1}\sin{u_1}-\cos{u_1}F_{12} +F_{13}=0;\quad \cr
F_{23,1}\sin{u_1} - F_{23}\cos{u_1} = 0;
\\\end{array}\ \right.\end{equation}

 \begin{equation}\label{47}
   \left\{\begin{array}{ll}

F_{12,3}+F_{23,1}=0;\quad F_{12,0}= F_{02,1};\quad \cr

F_{01,3}+ F_{13,0}=0;\quad F_{02,3}+ F_{23,0}=0.

 \\\end{array}\ \right.\end{equation}
From the system \eqref{45} we get:
$$F_{13}= a_1\sin{u_3}+b_1 \cos{u_3}, \quad F_{23}= -\sin{u_1}(a_1\cos{u_3}-b_1 \sin{u_3}), $$
$$F_{12}= \cos{u_1}(a_1\sin{u_3}+b_1 \cos{u_3})+c_{1}, $$
where \quad $a_1,\quad b_1 \quad c_1\quad $ functions of variables \quad $u_0,\quad u_1.$\quad Using the Bianchi identities \eqref{47} and rest equations of the system \quad \eqref{45}, \quad\eqref{46},\quad   we get:
$$a_1=\alpha_0 \quad b_1=\beta_0,\quad c_{1}=0. $$
Final solution can be present in the form:
\begin{equation}\label{48}
   \left\{\begin{array}{ll}
F_{01}= \dot{\gamma}_0 + (\dot{\alpha_0}\cos{u_3}-\dot{\beta_0} \sin{u_3}),\quad \cr
F_{02}= (\dot{\alpha_0}\sin{u_3}+\dot{\beta_0} \cos{u_3})\sin{u_1}, \cr
F_{03}=0, \cr
F_{12}= (\alpha_0\sin{u_3}+\beta_0 \cos{u_3})\cos{u_1},\quad \cr
F_{13}= \alpha_0\sin{u_3}+\beta_0 \cos{u_3}, \quad \cr
F_{23}= (-\alpha_0\cos{u_3}+\beta_0 \sin{u_3})\sin{u_1}.
\\\end{array}\ \right.\end{equation}

A particular solution of the system \eqref{26} in the selected gauge has the form:

\begin{equation}\label{49}
A_3=0, \quad
A_{1}=\gamma_0 + (\alpha_0\cos{u_3}-\beta_0 \sin{u_3}), \quad
A_{2}= (\alpha_0\sin{u_3}+\beta_0 \cos{u_3})\sin{u_1}.
\end{equation}

\section{Conclusion.}

In conclusion, we note some ways for using the obtained results.

1. The considered metrics define homogeneous spaces, due to which the results are of interest in cosmology, especially when studying the processes occurring in the early stages of the Universe  evolution.

2. The found external admissible fields, due to the special symmetry of homogeneous spaces, make it possible to construct interaction models of the axion field with the electromagnetic field, which is of interest when studying the problem of dark matter.

3. The results can be used to obtain exact self-consistent solutions in the General Theory of Relativity, in the scalar-tensor theory of gravity, in the Vaidya problem, as well as in the integration of field equations in other gravitational theories.

4. The results can be used to construct a theory of non-commutative integration of quantum motion equations in a strong gravitational field in the presence of fields of a gauge nature.

\section{Appendix}

For the sake of convenience, we present all the obtained results. For each group, the metric, electromagnetic potential, and integrals of motion are given.
The metrics were found in \quad \cite{38}. \quad
We follow the notation used in this book. All functions \quad $a_{\alpha\beta}$ \quad depend only on the variable \quad $u^0:$
$$a_{\alpha\beta}=a_{\alpha\beta}(u^0).$$

\subsection{Group $G_3(I).$}

1. Metrics:
$${ds}^2 =a_{\alpha\beta}du^\alpha du^\beta +e{du^0}^2,\quad e^2=1.$$
2. Potential of the admissible electromagnetic field:
$$A_0=0, \quad A_\alpha = \alpha_\alpha \quad \alpha_\alpha=\alpha_\alpha(u^0).\quad $$
3. Integrals of motion:
$$\hat{Y}_\alpha = \hat{p}_\alpha.$$

\subsection{Group $G_3(II).$}

1. Metrics:
$$ds^2 = {du^1}^2a_{11}+2{du^1du^2}(a_{12}+a_{11}u^3)+ 2{du^1du^3}a_{13}+ $$
 $${du^2}^2(a_{22}+2a_{12}u^3+a_{11}{u^3}^2)+2{du^2du^3}(a_{23}+a_{13}u^3)+$$
 $$+{du^3}^2a_{33}+e{du^3}^2, \quad e^2=1. $$

2. Potential:
$$
A_0=0,\quad A_1=\alpha_0,\quad A_2=\alpha_0 u_3+\beta_0, \quad A_3=\gamma_0.  .
$$

3. Integrals of motion:
$$
\hat{Y}_1 = \hat{p}_1, \quad \hat{Y}_2 = \hat{p}_2, \quad \hat{Y}_3 = u^2 \hat{p}_1 -  \hat{p}_3.
$$

\subsection{Group $G_3(III).$}

1. Metrics:\quad
$$
{ds}^2 = {du^1}^2a_{11}\exp{2u^3}+2{du^1du^2}a_{12}\exp{u^3}+ 2{du^1du^3}a_{13}\exp{u^3}+
$$
$$
2{du^2du^3}a_{23}+{du^2}^2a_{22}+{du^3}^2a_{33}+e{du^3}^2, \quad e^2=1.
$$

2. Potential of the admissible electromagnetic field:
$$
A_{0}=0, \quad A_{1}=\alpha_0\exp{u^3},\quad A_2=\beta_0,\quad A_3=\gamma_0.
$$

3. Integrals of motion:
$$
\hat{Y}_1 = \hat{p}_1, \quad \hat{Y}_2 = \hat{p}_2,
 \quad \hat{Y}_3 = u^1 \hat{p}_1 -  \hat{p}_3
$$

\subsection{Group $G_3(IV).$}

1. Metrics:\quad
$$
ds^2 = {du^1}^2a_{11}\exp{2u^3}+2{du^1du^2}(a_{12}+a_{11}u^3)\exp{2u^3}+ 2{du^1du^3}a_{13}\exp{u^3}+ $$
$$
2{du^2du^3}(a_{23}+a_{13}u^3)\exp{u^3}+{du^2}^2(a_{22}+2a_{12}u^3+a_{11}{u^3}^2)\exp{2u^3}
$$
 $$+{du^3}^2a_{33}+e{du^3}^2, \quad e^2=1.
$$

2. Potential of the admissible electromagnetic field:
$$
A_0=0,\quad A_1=\alpha_0\exp{u^3},\quad A_2 = (\alpha_0 u^3 + \beta_0)\exp{u^3},\quad A_3=\gamma_0.
$$

3. Integrals of motion:
$$
\hat{Y}_1 = \hat{p}_1, \quad \hat{Y}_2 = \hat{p}_2, \quad \hat{Y}_3 = (u^2 + u^1)\hat{p}_1 + u^2\hat{p}_2- \hat{p}_3.
$$

\subsection{Group $G_3(V).$}

1. Metrics:\quad
$$
ds^2 = {du^1}^2a_{11}\exp{2u^3}+2{du^1du^2}a_{12}\exp{2u^3}+ 2{du^1du^3}a_{13}\exp{u^3}+
$$
$$
2{du^2du^3}a_{23}u^3\exp{u^3}+{du^2}^2a_{22}\exp{2u^3}
$$
$$
 +{du^3}^2a_{33}+e{du^3}^2, \quad e^2=1.
$$

2. Potential of the admissible electromagnetic field:
$$ A_0=0, \quad A_1=\alpha_0\exp{u^3},\quad A_2 = \beta_0 \exp{u^3},\quad A_3=\gamma_0.$$
3. Integrals of motion:

$$
\hat{Y}_1 = \hat{p}_1, \quad \hat{Y}_2 = \hat{p}_2, \quad \hat{Y}_3 = u^1\hat{p}_1 + u^2\hat{p}_2- \hat{p}_3
$$

\subsection{Group $G_3(VI).$}

1. Metrics:
$$
{ds}^2 = {du^1}^2a_{11}\exp{2u^3}+2{du^1du^2}a_{12}\exp{3u^3}+ 2{du^1du^3}a_{13}\exp{u^3}+ $$
$$
 2{du^2du^3}a_{23}u^3\exp{2u^3}+{du^2}^2a_{22}\exp{4u^3}
$$
$$
+{du^3}^2a_{33}+e{du^3}^2, \quad e^2=1.
$$

2. Potential of the admissible electromagnetic field:
$$
 A_0=0, \quad A_1=\alpha_0\exp{u^3},\quad A_2 = \beta_0 \exp{2u^3},\quad A_3=\gamma_0.
$$

3.Integrals of motion:
$$
\hat{Y}_1 = \hat{p}_1, \quad \hat{Y}_2 = \hat{p}_2, \quad \hat{Y}_3 = u^1\hat{p}_1 + 2u^2\hat{p}_2- \hat{p}_3.
$$

\subsection{Group $G_3(VII).$}

1. Metrics:\quad
$$
{ds}^2 = {du^1}^2[a_{11}+a_{12}\cos{(2u^3 \sin{\alpha})}+a_{22}\sin{(2u^3 \sin{\alpha})}]\exp{(2u^3 \cos{\alpha})} +  2{du^1du^2}[a_{11}\cos{\alpha}+
(a_{12}\cos{\alpha}+
$$
$$
+a_{22}\sin{\alpha})\cos{(2u^3 \sin{\alpha})}+(a_{22}\cos{\alpha}-a_{12}\sin{\alpha})\sin{(2 u^3 \sin{\alpha})}]\exp{(2u^3 \cos{\alpha})}+
$$
$$
{du^2}^2[a_{11}+ (a_{12}\cos{2\alpha}+a_{22}\sin{2\alpha})\cos{(2u^3 \sin{\alpha})}+(a_{22}\cos{2\alpha}-a_{12}\sin{2\alpha})\sin{(2u^3 \sin{\alpha})}]\exp{(2u^3 \cos{\alpha})}
+$$
$$
2{du^1du^3}[(a_{13}\cos{\alpha}-a_{23}\sin{\alpha})\cos{(u^3 \sin{\alpha})}+ (a_{13}\sin{\alpha}+a_{22}\cos{\alpha})\sin{(u^3 \sin{\alpha})}] \exp{(u^3 \cos{\alpha})}+
$$
$$
2{du^2du^3}[a_{23}\sin{(2u^3 \sin{\alpha})}+a_{13}\cos{(2u^3 \sin{\alpha})}]\exp{(u^3 \cos{\alpha})}+{du^3}^2a_{33}+e{du^3}^2, \quad e^2=1.
$$

2. Potential of the admissible electromagnetic field:
$$
A_{0}=0, \quad A_{1}=(\alpha_0 \sin{(\alpha+u^3\sin{\alpha})} +\beta_0 \cos{(\alpha+u^3\cos{\alpha})})\exp{(-u^3 \cos{\alpha})},
$$

$$\quad A_{2}=(\alpha_0 \sin{(u^3\sin{\alpha})} +\beta_0 \cos{(u^3\cos{\alpha})})\exp{(-u^3 \cos{\alpha})},\quad A_3=\gamma_0.
$$

3. Integrals of motion:
$$\hat{Y}_1 = \hat{p}_1, \quad \hat{Y}_2 = \hat{p}_2, \quad \hat{Y}_3 =-u^2\hat{p}_1 +(2u^2 \cos{\alpha}+u^1)\hat{p}_2+\hat{p}_3.$$

\subsection{Group $G_3(VIII).$}

1. Metrics:\quad
$$
ds^2 = {du^1}^2a_{11}  +  2{du^1du^2}(a_{11}{u^1}^2-2a_{13}u^1 +a_{12})\exp{-u^3}+ 2du^1du^3(a_{13}-a_{11}u^1)+
$$
$${du^2}^2[a_{22}-4a_{23}u^1 +2(a_{12}+2a_{33}){u_1}^2-4a_{13}{u^1}^3+a_{11}{u^1}^4]\exp{-2u^3}
$$
$$ +2{du^2du^3}[a_{23}-(a_{12}+2a_{33})u_1+3a_{13}{u^1}^2-a_{11}{u^1}^3]\exp{-u^3}
$$
$$
+2{du^3}2(a_{11}{u^1}^2-2a_{13}u^1 +a_{33})+e{du^0}^2, \quad e^2=1.
$$

2. Potential of the admissible electromagnetic field:
$$
A_0=0, \quad A_{1}=\alpha_0,\quad A_{2}=(\alpha_0{u_1}^2 + 2\beta_0 u^1 +\gamma_0)\exp{-u^3},\quad A_3=-(\alpha_0{u^1} + \beta_0).
$$

3. Integrals of motion:
$$
\hat{Y}_1 = \hat{p}_2, \quad \hat{Y}_2 = u^2\hat{p}_2 +\hat{p}_3,\quad \hat{Y}_3 = \hat{p}_1 \exp{u^3} + \hat{p}_2 {u^2}^2 + 2u^2\hat{p}_3
$$

\subsection{Group $G_3(IX).$}

1. Metrics:
$$
{ds}^2 = {du^1}^2[a_{11}-(a_{12}\cos{2u^3}+a_{22}\sin{2u^3})]+2{du^1du^3}((a_{13}\cos{u^3}-
a_{23}\sin{u^3})+$$
$$
+2{du^1du^2}[(a_{13}\cos{u^3}-a_{23}\sin{u^3})\cos{u^1}+(a_{12}\cos{2u^3}-a_{22}\sin{2u^3})\sin{u^1}]
$$
$$
+{du^2}^2[a_{33}{\cos{u^1}}^2+(a_{23}\cos{u^3}+a_{13}\sin{u^3})\sin{2u^1}+
(a_{12}\sin{2u^3}+a_{22}\cos{2u^3}+a_{11}){\sin{u^1}}^2  ]
$$
$$
2{du^2du^3}(a_{33}\cos{u_1}+(a_{23}\cos{u^3}+a_{13}\sin{u^3})\sin{u^1})+{du^3}^2a_{33}+
e{du^0}^2.$$

2. Potential of the admissible electromagnetic field:
$$
A_0 = A_3=0, \quad
A_{1}= (\alpha_0\cos{u^3}-\beta_0 \sin{u^3}),
\quad A_{2}= (\alpha_0\sin{u^3}+\beta_0 \cos{u^3})\sin{u^1}.
$$

3. Integrals of motion:
$$
\hat{Y}_1 = \hat{p}_2, \quad\hat{Y}_2 = \hat{p}_1 {\cos{u^2}} + (\hat{p}_3 - \hat{p}_2 {\cos{u^1}})
\frac{\sin{u^2}}{\sin{u^1}},
\quad\hat{Y}_3 = -\hat{p}_1 {\sin{u^2}} + (\hat{p}_3 - \hat{p}_2 {\cos{u^1}})
\frac{\cos{u^2}}{\sin{u^1}}.
$$

 This work was supported by Ministry of Sceince and High Education of Russian Federation, project FEWF-2020-0003.


\end{document}